\documentstyle[12pt]{article}
\begin{document}

\title{A Method for  Justification of the View of Observables in
Quantum Mechanics and Probability Distributions in Phase Space}

\author{Evgeny Beniaminov\\
e-mail: beniamin@rsuh.ru\\
web site: http://www.rsuh.ru/win/whoiswho/2/HpageBen }
\date{}

\maketitle
\begin{abstract}
Let $f (x, p)$ be a function on the phase space.
The function $f$ corresponds to the classical observation.
Let $\rho (x, p,\xi) =\varphi^2 (x, p,\xi)$ be  some non-negative
density function
on an extended phase space. The density function $\rho$ corresponds
to the generalized state on the extended phase space. Then, as usual, the
observation $f$ in the generalized state $\rho$ is the value
$< f,\rho > = < f, \varphi^2 >$ of the integral of the function $f$ in
the distribution $\rho$. It is supposed, that in quantum
observations all distributions $\rho$ are realized not, and only
distributions $\rho=\tilde\varphi^2$, where $\tilde\varphi$ belongs
to some linear subspace $\tilde{\cal H}$ (averaging wave functions) in space of all functions on the extended phase space. Besides it is supposed, that
in quantum experiments values of the spectrum of the linear operator
$A_f$ of the quadratic form
$< f, \tilde\varphi^2 > = < \tilde\varphi, A_f \tilde\varphi^2 >$,
where $\tilde\varphi \in \tilde{\cal H}$, are observed.
In this paper we consider certain
hypotheses on the averaging process of wave functions.
Then it is shown that the spectrum of usual operator of the quantum
observable corresponded to $f$ is small differing from the spectrum of
the operator $A_f$. Other side, this approach enables one to recover
probability distributions in the phase space for  wave functions.
\end{abstract}
\section*{Introduction}
     In this paper there is considered a mathematical model of
microworld processes based on certain constructions of the geometric
quantization  \cite{K} which takes in account the  fluctuating affects
of the media, such as fluctuations of the vacuum.
We show that an enlargement of the phase space and of its motion
group and an account for the diffusion motions of microsystems in
the enlarged space, the motions which act by small random translations
along the enlarged group, lead to observable quantum effects.

     I show that certain conventional nonrelativistic descriptions
of  quantum  systems  can     be   obtained   as   asymptotic
approximations to the model suggested in this paper  with  respect
to the power series expansion in Planck's constant. The parameters
of  the  proposed  model  are  estimated  on  the  base   of   the
experimental data of Lamb's shifts in the spectrum  of  hydrogen's
atom \cite{LR}. In quantum mechanics these shifts can not be  explained
in the framework of the nonrelativistic quantum model and the idea
of  vacuum's  polarization  around  a  point  source  (radiational
corrections) is incorporated to do the trick.

    On the one hand, the suggested model can be  used  to  deepen
our perception on the nature of  quantum  effects;  on  the  other
hand, it leads to yet another  construction  of  quantizations  of
mechanical systems, the construction that might be of help in some
exceptional cases.

     Here I propose to  consider  probability  amplitudes  on  the
enlarged phase space which includes the space of inner  states  of
the system, its own in each  experiment.  I  assume  that  in  the
experiments on quantum systems  our  classical  gadgets  can  only
register  the  probability  distribution  corresponding   to   the
averaged probability amplitudes over fluctuating  actions  on  the
quantum system. I will  show  that  in  the  suggested  model  the
averaged  probability  amplitudes   on   the   phase   space   are
parametrized by complex functions on the configuration space, i.e.
the wave functions. In particular, formula~(\ref{tilde-rho})
of Teorem~4.1 obtained in  this
model reflects the probability distribution for a wave function in
the phase space. Wigner \cite{Wi} who obtained so-called
"quasidistributions" in the phase space was first solving this problem.
However in certain cases his quasidistributions can be negative
(unlike the ones obtained here) and therefore have no physical meaning.

     In sec.~1 the main hypotheses on quantum observations are
formulated and a formulation of the main problem is given.

     In sec.~2 the action of the enlarged group of motions on  the
enlarged phase space is described.

     In sec.~3 there are introduced hypotheses on  the  averaging
operation and the theorem on the form of averaged distributions in
the enlarged phase space is formulated.

     In sec.~4 the  main  result  of  the  paper  is  proved:  the
asymptotic proximity of linear operators of the  considered  model
to the linear operators of quantum observables and the probability
distribution in the phase space for a wave function.

     In sec.~5 there are  listed  several  unsolved  problems  and
trends of further investigations.

     Appendix contains  proof  of  the  theorem  on  the  form  of
averaged distributions in the enlarged phase space and an estimate
of the model's parameters based on the experimental data of Lamb's
shift in the spectrum of hydrogen's atom.

\section{Main hypotheses and formulation of the problem }
     First let us recall the main notions of classical mechanics for
the flat configuration space, cf. e.g. \cite{DNF,  B2}.  Let  $M$
(locally, $R^{2n}$) be the phase space with coordinates of  its
points being $(q,p)$, where $q=(q_1 ,...,q_n )$ is the point's
position  in  the configuration space and $p=(p_1 ,...,p_n )$
is its momentum.  In classical mechanics the dynamics is given
by the formula
                           $df/dt=\{f,H\},$
where $\{.,.\}$ is the Poisson  bracket  determined  by  the  exterior
2-form $\omega$ (locally of the form
$\sum_{1\leq i \leq n} dq_i \wedge dp_i$) and $H$ is a fixed function,
the Hamiltonian.

     The {\it canonical transformations} $g$ of the phase space
$M$  are (smooth, 1-1) maps $g: M \rightarrow  M$ preserving $\omega$.
Let $S$ denote the group of all canonical transformations
({\it symplectic}  diffeomorphisms)  of
the phase space. The dynamics of a mechanical system is given by a
1-parameter subgroup of $S$. To observables of a mechanical system
the functions $f$ on the phase space are assigned and to the  states
there are assigned nonnegative distributions $\rho dqdp$, or shortly, $\rho$.
The mean value of an observable $f$ in the state $\rho$ is by definition
           $$\langle f,\rho \rangle =\int_M f(q,p)\rho(q,p)dqdp.$$

We will make the following hypotheses on a quantum system  on
the above phase space:
\begin{enumerate}
     \item {\it The phase space is enlarged to the space $E$  which  is  the
total space of a fibration $\pi : E \rightarrow M$
 over $M$ with fiber $F$ which is a manifold
of 'inner states' of the system (if $M=R^{2n}$ then $E=R^{2n}\times F$).}
     \item {\it Let $pr: P\rightarrow S $ be the nontrivial central extension
of the group $S$ with the help of circle $T=R/hZ$ (Lie  algebra of $P$ is  the Poisson  algebra).
The group $P$  acts   by   diffeomorphisms on $E$  so   that
$\pi\circ g'=pr(g')\circ \pi$ for any $g'\in P$ where $g': E \rightarrow E$.}
     \item {\it The observables of a quantum  system,  as  well  as  of  a
classical one, are given by functions on the phase space.}
\end{enumerate}

     Denote by $\rho(q,p,\xi)dqdpd\xi$   a  nonnegative  distribution  on  E
determined by a distribution  density  $\rho(q,p,\xi)$  with  $\xi \in F$
 and  a $P$-invariant measure $dqdpd\xi$ on $E$. Let us present  $\rho$
in  the  form
$\rho(q,p,\xi)=\varphi^2(q,p,\xi)$,
where $\varphi(q,p,\xi)$ is a function on $E$ such that
                         $$\int_F  \varphi(q,p,\xi)d\xi=0.$$

Denote by $\tilde{\rho} (q,p,\xi)=\tilde{\varphi}^2 (q,p,\xi)$  an  averaged
density  distribution and $\tilde{\varphi} (q,p,\xi)$ obtained from
 $\varphi(q,p,\xi)$  by an averaging process under the action of small
fluctuations  of the system on $E$ (a  mathematical  model  of
this  averaging process will be given in sec.~3.

     $4'$. {\it Only averaged density distributions of the form
$\tilde{\rho} (q,p,\xi)= \tilde{\varphi}^2 (q,p,\xi)$
are realized in quantum measurements.
The map $\varphi \mapsto \tilde{\varphi}$  is  a  linear
projection operator whose form is  given  in  sec.3,  where  the
refined hypothesis 4), is given.}

     5. {\it Only averaged values of observables $f$ of the form
      $$\langle f,\tilde{\rho} \rangle =\int_E f(q,p)\tilde{\rho} (q,p,\xi)dqdpd\xi=
\int_E f(q,p)\tilde{\varphi}^2(q,p,\xi)dqdpd\xi $$
are realised in measurements, where
$\tilde{\rho} (q,p,\xi)= \tilde{\varphi}^2 (q,p,\xi)$ and $\tilde\varphi$ is the averaging function on $E$.}

     Now, consider the completed Hilbert space
$\tilde{\cal H} \subset L_2(E,dqdpd \xi)$
of averaged
functions
$\tilde{\varphi}(q,p,\xi)$ on $E$ with
 respect  to  the  standard inner product
 $$  \langle \tilde{\varphi}',\tilde{\varphi}'' \rangle  =
\int_E \tilde{\varphi}' \tilde{\varphi}'' dqdpd\xi.   $$

To a function $f$ on the phase space assign the linear operator  $A_f$
in $\tilde{\cal H}$  given by the formula
 $$   \langle \tilde{\varphi} ,A_f \tilde{\varphi} \rangle \stackrel{def}{=}
\langle f,\tilde{\rho} \rangle = \langle f,\tilde{\varphi}^2 \rangle.   $$

     The main result of this paper is the proof of the fact that
under the natural $P$-action on $E$ the operators $A_f$  are approximately
(up to terms of order $h$, where $h$ is  Planck's  constant)  coincide
with the operators of quantum observables in the accept definition
of quantum mechanics.

 \section{A description of the group $P$ and its action on
                   the enlarged space of states}
     Let $pr:P \longrightarrow S$ be the nontrivial central extension of the
group $S$ of canonical transformations with the help of the circle
(1-dimensional torus) $T=R/hZ$. Let ${\bf po}(2n)$ be the  Poisson  algebra,
the nontrivial central extension  of  the  Lie  algebra  ${\bf h}(2n)$  of
Hamiltonian  vector  fields.  In  a  sense  that  Lie  theory   is
applicable to infinite dimensional case ${\bf po}(2n)$ and ${\bf h}(2n)$
are  the Lie algebras of the groups $P$ and $S$, respectively.

 Two bundles $E,$ $E'$ over a symplectic manifold M  (with  fibers
$F,$ $F'$, respectively) with $P$-actions on  them  are  equivalent  if
there is a diffeomorphism of bundles with $P$-action compatible with
projections and the $S$-action on the base. To describe bundles with
$P$-action over $M$ we have to consider first  the  structure  of  the
fiber $F$ of the bundle $E$ in detail.

     Denote by $S_0$  the subgroup of  the  canonical  transformations $S$
that preserve the origin, let $P_0 =pr^{-1}(S_0 )$.
Clearly, $P_0 =T\oplus S_0.$

     Since $P_0 \subset P$ acts on $E$ and preserves the fiber  $F$ over  the
origin, a $P$-action in $F$ is defined. In particular, $T \subset P$  acts in
$F$.

 {\bf Theorem 2.1.} ([K]). {\it A bundle $E = R^{2n}\times F$
with  a $P$-action compatible with an ${\bf h}(2n)$-action on the space
of functions on $R^{2n}$ is uniquely up to equivalence of  $P$-bundles
determined  by the space $F$ with a $P_0$-action.}

  To a $P$-action on $E$ we assign in the standard way a Lie
algebra  homomorphism  ${\bf po}(2n)\longrightarrow {\bf vect}(E)$,
the  derivative  of  the
$P$-action on $E$. Denote by $\tau$ the vector field on $F$ corresponding  to
the action of the 1-parameter group generated by $T$  on  $E$  ($\tau$  can
also be defined as the value of the derivative of the action $T=T_t$ with
respect to the parameter $t$ at $t=0$).

     {\bf Corollary 2.2.} {\it  In  any  $P$-bundle  $E$  one   can
 chose   a trivialization (isomorphic to $R^{2n}\times F$) so that
the vector fields  on
$E$ corresponding to Hamiltonians linear in $q$ and  $p$,  i.e.  of  the
form $H= \sum_{i=1}^{n}(x_i p_i -y_i q_i)+c \in {\bf po}(2n)$, are  given
in  local  coordinates $(q,p,\xi)$ on $E$ by the expression
                  $$D_H =\sum_{i=1}^{n}(x_i \partial q_i +y_i \partial p_i)+
(c-\sum_{i=1}^{n} y_i q_i )\tau.$$
Equivalently, in the global form, this means that for an arbitrary
function $\varphi(q,p,\xi)$ on $E$ the action of the one-parametric  subgroups
$G^{H}_t$  for the above $H \in {\bf po}(2n)$ is given by the formula
 $$ G^{H}_{t} \varphi (q,p,\xi)=
\varphi (q+tx,p+ty,T_{t(c -\langle y, q \rangle )}(\xi)),$$
where  $\langle y, q \rangle = \sum_{i=1}^{n}y_i q_i$   and
$T_{t(c -\langle y, q \rangle )}$ is  the  element  of  the
1-parameter group $T$ corresponding  to  the  value
$t(c-\langle y,q \rangle )$  of the
parameter.}

In what follows we set:
                  $$ W_t^{x,y} = G_t^{H} \qquad  {\mbox for  }\ \
H= \sum_{i=1}^{n}(x_i p_i -y_i q_i).$$
Denote by $W$ the subgroup of $P$  generated  by
$W_t^{x,y}$ for $(x,y)\in R^{2n}$
(usually $W$ is called Heisenberg-Weyl group).  Clearly,  $W$  is  the
inverse image of $R^{2n}$  with  respect  to  the  projection
  $pr:P \rightarrow S$,
where $R^{2n} \subset S$ is considered as the subgroup of translations.

\section{Averaging}

     Now, let us pass to averaging of a function  $\varphi(q,p,\xi)$  on  a
$P$-bundle $E$. We need it in hypothesis 4.

     Let us start with the assumption that the  averaging
 $\varphi \mapsto \tilde{\varphi}$
is associated with a diffusion  process  (Brownian  motion)  on  a
$P$-bundle caused by  inaccuracy  of  the  setting  of  the  quantum
observational device. This process acts locally by small shifts by
$\Delta q$ and $\Delta p$ along the coordinates $q$ and $p$ of
$R^{2n}$, respectively, and globally with the help
of $W_1^{\Delta q,\Delta p}\in W $ acting on $E$.

     More precisely, let $\tau$ be the diffusion  time.  The  functions
are transformed as for a diffusion process, cf. e.g. [I]:
\begin{eqnarray} \label{intK}
\varphi(\tau + \Delta \tau, q, p, \xi)&=
&\int\limits_{R^{2n}} K(\Delta q, \Delta p, \Delta \tau)
\times\nonumber\\
& &\times
 W_1^{ \Delta q, \Delta p}\varphi(\tau, q, p, \xi)
d(\Delta q) d( \Delta p) + o(\Delta \tau),
\end{eqnarray}
where $ K(\Delta q, \Delta p, \Delta \tau) $ is the probability
density  of  shifts  by  the
vector \linebreak $ (\Delta q, \Delta p)$  during the time $\Delta \tau$.

     Naturally, we make the usual assumptions about
$ K(\Delta q, \Delta p, \Delta \tau)$:

---$ K(\Delta q, \Delta p, \Delta \tau) $ is rapidly decreasing at infinity;

--- the mathematical expectation of the shift vector
$(\Delta q, \Delta p) $  is zero;

--- the diagonal elements of the matrix of 2nd moments  are  of
the form
$$
\int_{R^{2n}} (\Delta q_i)^2 K(\Delta q, \Delta p, \Delta \tau)
d(\Delta q) d(\Delta p) =
2a_i^2 \Delta \tau + o(\Delta \tau),
$$
$$
\int_{R^{2n}} (\Delta p_i)^2 K(\Delta q, \Delta p, \Delta \tau)
d(\Delta q) d(\Delta p) =
2b_i^2 \Delta \tau + o(\Delta \tau),
$$
where $a$  and $b$ characterize  the  "intensity"  of  shifts  along
positions and momentum;

     --- shifts along distinct directions are poorly correlated with
each other, i.e. the offdiagonal 2-nd moments are of order $o(\Delta \tau)$;

--- the moments of $ K(\Delta q, \Delta p, \Delta \tau)$ of orders
greater  than  2  are also of order $o(\Delta \tau).$

     Starting from~(\ref {intK})  and  the  above  assumptions  we  expand
$\varphi(\tau, q, p, \xi)$ in  the  Taylor  series  to  derive
as $\Delta \tau \rightarrow 0$  a differential equation  similar
to  the  diffusion  equation.  The equation  represents  the  averaging
of  the  function  $\varphi$   over
fluctuations. The asymptotic of solutions  of  this  equation  as
$ \tau \rightarrow \infty$ is given by the following theorem,
where $j$ is the  imaginary
unit, $*$ denotes the complex conjugation and $\psi(x, \xi)$ is
a  complex-valued function on $R^{n}\times F$ such that
\begin{equation}\label{psi_t}
\psi(x, T_{t}(\xi)) =\psi(x, \xi) \exp \left( -j\frac{2\pi}{h}t \right).
\end{equation}

 {\bf T h e o r e m 3.1.}{\it Let $\varphi(\tau, q, p, \xi)$
satisfy~(\ref{intK}) and $\varphi(0, q, p, \xi) = \varphi(q, p, \xi)$,
where
                          $\int_0^h \varphi(q, p, T_t(\xi)) dt = 0.$
     Then $\varphi(\tau, q, p, \xi)$ asymptotically approximates,
as $ \tau \rightarrow \infty$, to the function
\begin{equation}
\tilde{\varphi}(q, p, \xi) \exp \left( -\tau \sum^n_{i =1}
\frac{2\pi a_i b_i}{h} \right),
\end{equation}
where
\begin{eqnarray}\label{tilde-phi}
\tilde{\varphi}(q, p, \xi)\!\!\!&=
&\!\!\!\frac{1}{\sqrt {2}}\left( \frac{2}{h^3} \right)^{\frac{n}{4}}
\left( \frac{b_1 ... b_n}{a_1 ... a_n}\right)^{\frac{1}{4}}
\int_{R^n} \exp\! \left(-\frac{\pi }{h }\sum_{i=1}^n\frac{b_i}{a_i}
 \left( q_i - x_i \right)^{2} \right)\times \nonumber\\
&\times &\!\!\!
\left(
\psi (x, \xi) \exp\! \left(-j\frac{2\pi \langle p, x \rangle}{h}\right) \! +
\psi^{*} (x, \xi) \exp\! \left(j\frac{2\pi \langle p, x \rangle}{h}\right)
\right)\! dx.
\end{eqnarray}
The  function  $\psi(x, \xi)$  is  obtained  from  the  function
$\varphi(q, p, \xi)$
according to the formula
\begin{eqnarray}\label{psi}
\psi(x, \xi)\!\! & = &\!\!\sqrt {2}\left( \frac{1}{h} \right)
\left( \frac{2}{h^3} \right)^{\frac{n}{4}}
\left( \frac{b_1 ... b_n}{a_1 ... a_n}\right)^{\frac{1}{4}}
\int_0^h \int_{R^{2n}} \varphi \left(q, p, T_{t}(\xi)\right) \exp\!
\left(j\frac{2\pi t}{h}\right)\times \nonumber\\
& &\!\!
\times\! \exp\! \left(-\frac{\pi}{h}\sum_{i=1}^n
\frac{b_i}{a_i} \left( q_i - x_i \right)^{2} \right)
\!\exp\! \left(j \frac{2\pi \langle p, x \rangle}{h}\right)
dq dp dt.
\end{eqnarray}
Besides,  if  $\psi(x, \xi)$ is  an  arbitrary  complex-valued   function
satisfying~(\ref{psi_t}) then the composition of transformations
$\psi  \mapsto \tilde{\varphi}$ and $\tilde{\varphi} \mapsto \psi$
 given  by
(\ref{tilde-phi}) and (\ref{psi}) is identity.}

     Proof is given in Appendix.

 Now we are ready to refine hypothesis~$4'$ from sec.~2 as follows:

     4. {\it Let $\varphi(q,p,\xi)$ be a real-valued  function  on
$E$ such  that $$\int_0^h \varphi(q,p,T_t(\xi))dt=0$$ and
$\rho(q,p,\xi)=\varphi^2(q,p,\xi)$ be the probability  density
on $E$. The averaging operation mentioned in hypothesis~$4'$ from section~2 is
caused by the fluctuation  process  described  by  equation~(\ref{intK})
where $\varphi$ is the initial state and $\tilde{\varphi}$  is
the asymptotic one  as  the time of fluctuation tends to infinity.}

     A corollary of this hypothesis: the map
$\varphi\mapsto\tilde{\varphi} $  is  obtained
(by Theorem~3.1) as the composition  of  maps  given  by
formulas~(\ref{psi})
 and (\ref{tilde-phi}) respectively.

 Consider the Hilbert space $\tilde{\cal H}$ of real functions
$\tilde{\varphi}$ on  $E$  of
the form (\ref{tilde-phi}) and with the standard inner product
 $$ \langle \tilde{\varphi}', \tilde{\varphi}'' \rangle =
 \int_E \tilde{\varphi}' \tilde{\varphi}'' dqdpd\xi.   $$
Denote by ${\cal H}$ the Hilbert space of complex-valued  functions
$\psi(x,\xi)$
on $R^n \times F$ satisfying~(\ref{psi_t}) and with the inner product
given  by  the formula
$$  \langle \psi_1 , \psi_2 \rangle=
Re \int_{R^n \times F} \psi_1(x, \xi) \psi_2^{*}(x, \xi) dxd\xi.$$

     Making use of Theorem 3.1 we directly derive the following

     {\bf C o r o l la r y 3.2.} {\it The map given by formula
(\ref{tilde-phi}) is  an  isomorphism of Hilbert spaces ${\cal H}$
and $\tilde{\cal H}$.}

     A function $\psi (x, \xi)$ satisfying (\ref{psi_t})  will  be
called  a  wave function, the corresponding function $\tilde\varphi(q,p,\xi)$
will be  called  the probability amplitude in the enlarged phase space.

     {\bf R e m a r k.}  Formula (\ref{psi})   gives   the   so-called
position representation of $\varphi$. The momentum representation of $\varphi$
is given  by the Fourier transform of $\psi (x,\xi)$ with respect to  $x$.  We
will  not n¥ed it.

\section {An estimate of the operator of an observable }

     In this section we will estimate the operator  ${\tilde A}_f$   acting  in
the Hilbert space $\tilde {\cal H}$ of functions of the form~(\ref{tilde-phi}).
By  definition ${\tilde A}_f$  is given by the following expression:
\begin{equation} \label{defA}
      \langle f, \tilde{\rho} \rangle_{\tilde {\cal H}} =
\langle \tilde{\varphi},A_f \tilde{\varphi} \rangle_{\tilde {\cal H}}  =
\int_E f(q,p) \tilde{\varphi}^2(q,p,\xi)dqdpd\xi.
\end{equation}

Since by Corollary 3.2 the Hilbert spaces $\tilde{\cal H}$ and ${\cal H}$ are
isomorphic, we can estimate $A_f$  by  estimating  image of ${\tilde A}_f$ in
${\cal H}$   under  this isomorphism.

     Let us substitute (\ref{tilde-phi}) in (\ref{defA}). Represent
the function
 $$\tilde{\varphi}^2 (q, p, \xi) =
\left( \int_{R^n} B(q, p, \xi, x)dx \right)^2,$$
where $B(q, p, \xi, x)$ is the integrand of (\ref{tilde-phi}),
in the form
  $$ \tilde{\varphi}^2 (q, p, \xi)=
\int_{R^n}\int_{R^n}  B(q,p,\xi,x')B(q,p,\xi,x)dx'dx. $$

Since the integrals  of  the  product  $\psi(x',\xi) \psi(x,\xi)$  and
of the product
$\psi^*(x',\xi) \psi^*(x,\xi)$  over $\xi$  are  zero  due  to~(\ref{psi_t}),
we  get  after simplification
\begin{eqnarray}
\langle f, \tilde{\rho} \rangle_{\tilde {\cal H}} &=&
\langle \tilde{\varphi},A_f \tilde{\varphi} \rangle_{\tilde {\cal H}}=
{\langle \psi, A_f \psi \rangle}_{\cal H}
=\left(\frac{2}{h^3}\right)^{\frac{n}{2}}
\left(\frac{b_1 ... b_n} {a_1 ... a_n} \right)^{\frac{1}{2}} \times\nonumber\\
& &\times \int \limits_{R^n} \int \limits_{R^n} \int \limits_{E}
    f(q,p)\exp\left[ -\frac{\pi}{h}\sum_{i=1}^{n}
    \frac{b}{a}((q_i -x'_i)^2 +(q_i -x_i)^2)\right] \times\nonumber\\
& &\times \psi(x', \xi) \psi^{*}(x, \xi)
\exp \left(-j\frac{2\pi \langle p,x-x' \rangle}{h}\right)dqdpd\xi dx'dx. \nonumber
\end{eqnarray}
This implies that the kernel of $A_f$  on the space ${\cal H}$ is of the form
\begin{eqnarray}\label{A_f}
 A_f (x,x')
&=&\left(\frac{2}{h^3}\right)^{\frac{n}{2}}
\left(\frac{b_1 ... b_n} {a_1 ... a_n} \right)^{\frac{1}{2}} \times\nonumber\\
& &\times \int \limits_{R^{2n}}
    f(q,p)\exp\left[ -\frac{\pi}{h}\sum_{i=1}^{n}
    \frac{b}{a}((q_i -x'_i)^2 +(q_i -x_i)^2)\right] \times\nonumber\\
& &\times \exp \left(-j\frac{2\pi \langle p,x-x' \rangle}{h}\right)dqdp,
\end{eqnarray}
and the density of the  probability  distribution
$\tilde{\rho} (q,p)={\tilde{\varphi}}^2(q,p)$
corresponding to $\psi(x,\xi)$ is given by the next theorem.

{\bf T h e o r e m 4.1.} {\it Let $\psi(x,\xi)$ be a wave function (in the position
representation) on $R^n \times F $ satisfying  (\ref{psi_t}); then the density
of the density of the  probability  distribution on the phase space is given
by the formula:}
\begin{eqnarray}\label{tilde-rho}
 \tilde{\rho} (q,p)
&=&\left(\frac{2}{h^3}\right)^{\frac{n}{2}}
\left(\frac{b_1 ... b_n} {a_1 ... a_n} \right)^{\frac{1}{2}} \times\nonumber\\
& &\times \int \limits_{F} \int \limits_{R^{n}} \int \limits_{R^{n}}
    \exp\left[ -\frac{\pi}{h}\sum_{i=1}^{n}
    \frac{b}{a}((q_i -x'_i)^2 +(q_i -x_i)^2)\right] \times\nonumber\\
& &\times \exp \left(-j\frac{2\pi \langle p,x-x' \rangle}{h}\right)
\psi(x', \xi) \psi^{*}(x, \xi) dx' dx d\xi.
\end{eqnarray}

This distributions is different from Wigner's quasidistributions by the integration  with the function
$$\left(\frac{2}{h}\right)^{\frac{n}{2}}
\left(\frac{b_1 ... b_n} {a_1 ... a_n} \right)^{\frac{1}{2}}
 \exp\left[ -\frac{\pi}{h}\sum_{i=1}^{n}
    \frac{b}{a}((q_i -x'_i)^2 +(q_i -x_i)^2)\right]. $$

Now in order to represent the operator $A_f$  in the form habitual in
quantum mechanics (cf. \cite{FYa}), substitute in~(\ref{A_f})  the  expression
of $f$ in terms of its Fourier transform:
      $$ f(q,p)=\left(\frac{1}{2\pi}\right)^n \int \limits_{R^{2n}}
\hat {f}(u,v)\exp[-j(\langle q,v \rangle + \langle p,u \rangle )]dudv,$$
where
$$ \hat {f}(q,p)=\left(\frac{1}{2\pi}\right)^n \int \limits_{R^{2n}}
f(u,v)\exp[j(\langle q,v \rangle + \langle p,u \rangle )]dqdp.$$

In the obtained formula we first integrate over $q$, then over $p$ and
$u$. We get:
\begin{eqnarray}
A_{f}(x, x') &=&\frac{1}{h^n} \int\limits_{R^{2n}} {\hat f} \left(\frac{2\pi
(x-x')}{h}, v \right)
\exp \Biggl(-\frac{h}{8\pi} \sum_{i=1}^n
\Biggl(\frac{a_i}{b_i}v^2_i +\nonumber\\
& &+\frac{b_i}{a_i} \Biggl(\frac{2\pi(x_i-x'_i)}{h} \Biggr)^2
\Biggr) \Biggr)
\exp \left(-j\left\langle \frac{x+x'}{2}, v \right\rangle \right) dv =
 \nonumber\\
&=&\frac{1}{h^n} \int\limits_{R^{2n}}\tilde f_h \left(\frac{2\pi (x-x')}{h}, v \right)
\exp\!\left(-j\left\langle \frac{x+x'}{2}, v \right\rangle \right) dv , \nonumber
\end{eqnarray}
where
$$ \tilde f_h(u, v) = \hat f (u, v) \exp \left(- \frac{h}{8\pi}\sum_{i=1}^n \left(\frac{a_i}{b_i}v_i^2 +
\frac{b_i}{a_i}u_i^2 \right)\right).
$$

     If instead of $\tilde f_h (u,v)$ we take its Taylor series expansion  in
powers of $h$ up to order n  we  get  the  corresponding  asymptotic
representation of $A_f$. In particular, the asymptotics  of  the  0-th
term of the expansion is
$$
A_{f}(x, x') =  \frac{1}{h^n} \int_{R^{n}}\hat f \left(\frac{2\pi (x-x')}{h}, v \right)
\exp \left(-j\left\langle \frac{x+x'}{2}, v \right\rangle \right) dv.
$$

     This expression coincides with the expression for the
operator of an observable in quantum mechanics  (formula  (14)  in
[FYa] whose $h$ is our $h/2\pi$). This implies the next  result  of  the
paper:

    {\bf T h e o r e m 4.2.} {\it The linear operator $A_f$  of  the
 form~(\ref{A_f})    in  the
Hilbert space ${\cal H}$ which is constructed from the classical observable
$f$ under assumptions 1-5 on the process of quantum  observation  is
asymptotically close  to  the  conventional  operator  of  quantum
observable given in the position representation.}

  If a more precise estimate of $A_f$  is required  we  can  always
take into account more terms of the  Taylor  series  expansion  of
$f_h (u,v)$ with respect to $h$.

     Examples of exact formulas for $A_f$.  By the  usual  abuse  of
language let  us  denote  the  operator  of  multiplication  by  a
function $F$ by $F$. We have:
\begin{eqnarray*}
A_{q_i}&=&x_i;\\
A_{p_i}&=&-j\frac{h}{2\pi}\frac{\partial}{\partial x_i};\\
A_{q_i^2}&=&x_i^2 + \frac{h a_i}{4\pi b_i};\\
A_{p_i^2}&=&- \frac{h^2}{4\pi^2}\frac{\partial^2}{\partial x_i^2} +
\frac{h b_i}{4\pi a_i}.
\end{eqnarray*}

It follows that for the Hamiltonian
$f = p^2/({2m})+ {m\omega^2 q^2}/{2}$
of the linear oscillator with eigen frequency $\omega$ we have
$$ A_f = -\frac{h^2}{8\pi^2m}\frac {\partial^2}{\partial x^2} + \frac{m\omega^2 x^2}{2}+ \sum_{i=1}^3
\frac{h(b_i^2 + m^2 \omega^2 a_i^2)}{8\pi a_i b_i m}. $$
which differs from the conventional operator   of the quantum linear oscillator
 by constants (the last summand).

In other side, if $f=f(q)$, i.e. does not depend  on  momenta,
we deduce from~(\ref{A_f}) after integration over $p$ and $x'$  that  $A_f$
is the operator of multiplication by
$$\bar f(x) =
\left( \frac{2}{h} \right)^{\frac{n}{2}}
\left( \frac{b_1 ... b_n}{a_1 ... a_n}\right)^{\frac{1}{2}}
\int\limits_{R^{n}}f(q) \exp \left(-\frac{2\pi}{h}
\sum_{i=1}^n \frac{b_i}{a_i} ( q_i - x_i )^2 \right) dq.
$$

The passage $f \mapsto \bar f$  is the
convolution of $f$  with
the density of the probability distribution with dispersion  along
the $q_i$-axis equal to ${h a_i}/({4\pi b_i}).$

     Suppose
$$ a_1 /b_1 =a_2 /b_2 =a_3 /b_3 =a/b.$$
Then for the Hamiltonian
$$ f(q, p) =({p_1^2 + p_2^2 + p_3^2})/({2m}) + V(q_1, q_2, q_3), $$
we have
\begin{equation}\label{A_f_hidr}
A_f = - \frac{h^2}{8\pi^2m}\left(\frac {\partial^2}{\partial x_1^2} +
\frac {\partial^2}{\partial x_2^2} + \frac {\partial^2}{\partial x_3^2}   \right) +
\frac{3h b}{4\pi a} + \bar V(x_1, x_2, x_3).
\end {equation}

     In particular for the Hamiltonian of the hydrogen atom whose
\begin{equation}\label{V_hidr}
V(q_1, q_2, q_3) =-e^2/r,
\end{equation}
where $e$ is the charge of the electron and $r^2 =\sum_{i=1}^3 q_i^2$,
  the  operator
$A_f$  differs from the operator by an irrelevant constant
${3h b}/({4\pi a})$ and
the extra smoothness of the Coulomb potential. Thus our hypotheses
predict  that  the  theoretical  spectrum  of  the  hydrogen  atom
computed on the base of the conventional Hamiltonian should differ
from that obtained in experiments.
     Such a  descrepancy  of  theory  and  experiment  was  indeed
detected in 40s [LR]. It is called Lamb's shift of hydrogen atom's
levels and is conventionally explained in quantum  electrodynamics
by  an  interaction   of   the   election   with   a   fluctuating
electromagnetic field.
     Comparison of these experimental data  with  calculations  of
the spectrum of $A_f$  given by~(\ref{A_f_hidr}), (\ref{V_hidr})
via  perturbation  theory
yields the following estimate of parameters of our model  (details
see in Application A.2):
$$a/b = 3,41\cdot 10^4 sec/gr;\ \ \ \ \
\Delta q =\sqrt{\frac{h a}{4\pi b}}=4,24\cdot 10^{-12} cm.$$

Hence  the  standard  deviation   of   the   normal   distribution
$\Delta q$ is comparable with
${h}/({2\pi m c}) = 3,8\cdot  10^{-11} cm, $
the minimal position
error  of  electron  (in  the  rest  frame)  obtained  in  quantum
electrodynamics.

\section{Conclusion and plan of further study}

     The description of quantum  systems  is  usually  based  upon
certain formal procedures starting from a classical description of
the  corresponding  mechanical  systems. Many physicists and mathematicians,
starting with Einstein, searched for a meaning of these procedures
but the success of quantum mechanics approved the formal  approach
to the quantization procedure and von Neumann's theorem on "hidden
parameters" [Nu] discredited for a long time such a  search  as  a
direction of scientific investigations.

The interest to the problem was revived in 50s  in  works  of
Bohm and de Broglie [B], [Br] and maintained in a number of  later
works, cf. [BV], [N], [PG], [M], [Ba], [KV]. At the same time  the
rigidity and lack of motivation in the  mathematical  requirements
in von Neumann's theorem became manifest [F].

     A detailed analysis of the problem of introduction of  hidden
parameters in quantum mechanics and von Neumann's theorem is given
in [Kh]. In particular, there is given a formal model with  hidden
parameters for a "solitary"  quantum  system.  The  difficulty  of
introducing  the  classical  probabilistic   model   for   quantum
phenomena is  associated  with  a  non-local  character  of  these
phenomena, confirmed in a number of experiments,  cf.  the  review
[Gr], [SM] (however, cf. [B1], Appendix~1).

     We have shown how to deduce operators of quantum  observables
on  the  base  of  hypotheses  of  sec.~1.  The  reason   causing
fluctuations  on  the  extended  phase  space   are   diverse:   a
fluctuating external force, inaccuracy in the description  of  the
real system, etc. We have shown that irrespective of the nature of
the fluctuation quantum effects will be observed in  such  systems
with accuracy determined by the Planck's constant and  the  ratios
$a_i /b_i$, where $a_i$  and $b_i$  are the intensities of
random shifts  along
the $i$-th position and momentum, respectively.

     Elsewhere I intend to investigate the following problems:

     --- Take into  the  account  relativistic  effects.   We   have
constructed  $A_f$   by  formula~(\ref{A_f})  having  given   a   classical
observable $f$ under the assumptions that the intensities of  shifts
a  and b  along the positions and momenta are constants.  This  is
not a relativistic requirement and it is  desirable  to  find  the
dependence of  the  intensities  on  momenta  and  the  masses  of
particles and to refine formula~(\ref{tilde-phi}).

      --- Take into the account spin. For this we should replace  our
group ${P}$ by a supergroup whose Lie superalgebra is
${\bf po}(2n|m)$ or  its
"odd" counterpart ${\bf b}_{\lambda} (n)$, see [L].

--- Generalize our construction to phase  spaces  more  general
than direct products of the configuration space by  the  space  of
momenta. This should lead to restrictions on the Planck's constant
$h$ cases by the geometry of the phase space, cf. [B2].

--- Describe  dynamics  of  observable  taking  into   account
fluctuating action of the ambient media. As is shown in App.1  the
time for stabilization (averageing) over fluctuations is of  order
${h}/({2\pi a b})$.
If  this  quantity  is  small   the   dynamics   can   be
approximately described as the superposition of a fast and a  slow
movements. The fast motion  leads  to  averageing  of  probability
amplitudes and the slow one describes the motion in the  space  of
averaged amplitudes.  The  classical  Schroedinger  equation  only
describes the slow constituent of the motion of a microobject.

\bigskip
\begin{center} {\Large Appendix 1. Proof of Theorem 3.1}
\end{center}
     Since (3.1)  is   determined   via   the   action   of   the
Heisenberg-Weyl group $W_t^{x, y}$ in the space  of  square
integrable  (and complexified for convenience)
functions on $E$, let us  decompose  a
function  $\varphi(\tau, q, p, \xi)$    into   the   integral   over   irreducible representations of $W$.

     The decomposition will be performed into three steps.
     First, let us consider the action of $T=R/hZ$ in the  space  of
functions on $F$. Let $\varphi(\xi)$ be a function on $F$. Since
$\varphi(T_t (x))$, where
$T_t\in T$, is periodic in $t$ with period $h$, it has  the
Fourier  series expansion
$$\varphi(T_t(\xi)) =\sum_{k\in Z} \varphi_k(\xi) \exp
\left(j\frac{2\pi k t}{h}\right),$$
where
$$\varphi_k (\xi) = \frac{1}{h}\int_t^h  \varphi(T_t(\xi)) \exp
\left(-j\frac{2\pi kt}{h}\right) dt.  $$

In particular, for $t=0$ we get
$\varphi(\xi)=\sum_{k\in Z} \varphi_k(\xi).$
 It follows  from  the
definition of $\varphi_k(\xi)$  that $\varphi(\xi) \mapsto \varphi_k(\xi)$
  is a projection and  $T$  acts  on  the
functions $\varphi_k(\xi)$  by the formula
$$ \varphi_k (T_t(\xi)) = \exp\left(j\frac{2\pi kt}{h}\right) \varphi_k(\xi). \eqno   (A.1)
$$

As is easy to see, the functions $\varphi_k$  and $\varphi_n$  are orthogonal for $n\ne k$.

     A function $\varphi(q, p, \xi)$ on  the  $P$-bundle  $E=R^{2n}\times F$
can  also  be represented in the form
$$ \varphi(q, p, \xi) = \sum_{k\in Z} \varphi_k(q, p, \xi), \eqno  (A.2)
$$
where
$$\varphi_k(q, p, \xi) = \frac{1}{h}\int_t^h  \varphi(q, p, T_t(\xi))
\exp\left(-j\frac{2\pi kt}{h}\right)  dt.  $$
Having substituted (A.2) into (3.1) and taking (A.1) into account
we  get  the  following  equations  for  each  of  the  orthogonal
components of $\varphi$ in coordinates on $E$ given by Corollary~2.2
$$
\varphi_k(\tau + \Delta \tau, q, p, \xi) = \int_{R^{2n}} K(\Delta q, \Delta p, \Delta \tau)
\varphi_k(\tau, q+\Delta q, p+\Delta p, \xi)
$$
$$
\times\exp\left(-j\frac{2\pi k\langle \Delta p, q \rangle}{h}\right)
d(\Delta q) d( \Delta p) + o(\Delta \tau).
 \eqno  (A.3)
$$

     For $k=0$ the equation (A.3) turns as $\Delta\tau \to 0$
(as is clear  from
the Taylor  series  expansion)  into  the  following  differential
equation on $\varphi_0(\tau, q+\Delta q, p+\Delta p,\xi)$:
$$
\frac{\partial\varphi_0}{\partial\tau} = \sum_{i=1}^n \left( a_i^2\frac{\partial^2\varphi_0}{\partial q_i^2} +
b_i^2\frac{\partial^2\varphi_0}{\partial p_i^2}\right),
$$
and since by hypothesis of Theorem 3.1
$$
\varphi_0(0, q, p, \xi) = {1}/{h}\int_t^h  \varphi(q, p, T_t(\xi)) dt=0,
$$
then $ \varphi_0(\tau, q, p, \xi) \equiv 0$ for any $\tau \ge 0$.

     For $k\ne 0$ let us expand $\varphi_k(\tau, q, p, \xi)$ into the
Fourier  integral
with respect to $p$, i.e. represent it in the form:
$$
\varphi_k(\tau, q, p, \xi) = \left( \frac{|k|}{h }\right)^{\frac{n}{2}}
\int_{R^n} \hat{\varphi}_k(\tau, q, x, \xi)
\exp\left(j\frac{2\pi k \langle p, x \rangle}{h}\right) dx, \eqno   (A.4)
$$
where
$$
\hat{\varphi}_k(\tau, q, x, \xi) = \left( \frac{|k|}{h}\right)^{\frac{n}{2}}
\int_{R^n} \varphi_k(\tau, q, p, \xi)
\exp\left(-j\frac{2\pi k \langle p, x \rangle}{h}\right) dp. \eqno (A.5)
$$

     {\bf R e m a r k.} As is easy to verify, the functions of the form
$$\hat{\varphi}_k(\tau, q, x, \xi) \exp
\left(j {2\pi k \langle p, x \rangle}/{h}\right)
\ \ \ \mbox{ for fixed }\ \ k,\ x,\ \xi   \eqno (A.6)$$
span a $W$-invariant subspace in the space of functions  on  $E$  with
the $W$-action described in the Corollary 2.2.

     Since the spaces of functions (A.6) are orthogonal to each
other for distinct $k, x$ and are $W$-invariant, then  equations  (A.3)
split into a system of equations, each  for  each  subspace,  each
obtained from (A.3) by substituting $\hat{\varphi}_k(\tau, q, x, \xi) \exp\left(j {2\pi k \langle p, x \rangle}/{h}\right)$.

     As a result of all these transformations we get
\begin{eqnarray*}
\hat{\varphi}_k(\tau + \Delta \tau, q, x, \xi) &=& \int_{R^{2n}} K(\Delta q, \Delta p, \Delta \tau)
\hat{\varphi}_k(\tau, q+\Delta q, x, \xi)\times\\
&\times &\exp\left(j\frac{2\pi k\langle \Delta p, x-q \rangle}{h}\right)
d(\Delta q) d( \Delta p) + o(\Delta \tau),
\end{eqnarray*}

which by expansion of the function
$$\hat{\varphi}_k(\tau, q+\Delta q, x, \xi) \exp\left(j {2\pi k \langle \Delta p, x-q \rangle}/{h}\right) $$
into the Taylor series in powers of $\Delta p$ and $\Delta q$
 and with  the  above
properties of  $K(\Delta q, \Delta p, \Delta \tau)$
 take as $\Delta \tau \to 0$ the form
$$
\frac{\partial\hat{\varphi}_k}{\partial\tau} =\sum_{i=1}^n \left(
a_i^2\frac{\partial^2\hat{\varphi}_k}{\partial q_i^2} +
b_i^2\left(\frac{2\pi k}{h}(x_i-q_i)\right)^2\hat{\varphi}_k \right), \eqno    (A.7)
$$
where $\hat{\varphi}_k$  are functions in $\tau, q, x, \xi$ for
$k\in Z\setminus \{0\}$.

     To study eqs. (A.7), consider  the  corresponding  eigenvalue
problems:
$$
a_i^2\frac{\partial^2\hat{\varphi}_k}{\partial q_i^2} +
b_i^2\left(\frac{2\pi k}{h}(x_i-q_i)\right)^2\hat{\varphi}_k =
\lambda \hat{\varphi}_k.
$$
This  is  a  stationary   Schroedinger   equation   for   harmonic
oscillations [FYa]. If $\hat{\varphi}_k(q) \rightarrow 0 $   as $|q| \rightarrow
\infty,$ the  equation  has  a
discrete spectrum. Its eigen values are of the form
$$
\lambda_{k, k_1, ..., k_n} = - \sum_{i=1}^n \frac{2\pi |k| a_i b_i}{h}(2k_i+1),
\eqno   (A.8)
$$
 where £¤¥ $k_1, ... ,  k_n$  are nonnegative integers. The corresponding  eigen
functions $ \hat{\varphi}_{k, k_1, ..., k_n}(q, x)$ are products of  the  Chebyshev-Hermit
polynomials        in      $({b_i}/{a_i})\sqrt{{2\pi |k|}/{h}}(q_i - x_i)$
 by
$\exp\left( -{\pi |k|}/{h}\sum_{i=1}^n  ({b_i}/{a_i})(q_i - x_i)^2 \right).$

     Since  eigen  functions  of  the  Schroedinger  equation  are
orthogonal to each other and constitute a complete system  in  the
space of  square  integrable  functions  in  $q$,  then $ \hat{\varphi}_k$
  can  be
represented as the series:
$$
 \hat{\varphi}_k (\tau, q, x, \xi) = \frac{1}{\sqrt 2}\sum_{ k_1, ..., k_n =0}^\infty
 c_{k, k_1, ..., k_n}(\tau, x, \xi)\  \hat{\varphi}_{k, k_1, ..., k_n}(q, x), \eqno  (A.9)
$$
where
$$
c_{k, k_1, ..., k_n}(\tau, x, \xi) =\sqrt{2} \int\limits_{R^n}
\hat{\varphi}_k (\tau, q, x, \xi) \ \hat{\varphi}_{k, k_1, ..., k_n}(q, x) dq.
$$

The multiple $1/\sqrt 2$ is taken for convenience.
     In particular, the maximal eigen value
$\lambda_{\pm 1, 0, ... , 0} = -\sum_{i=1}^n 2\pi a_i b_i/h$
is attained on the normed eigen function
$$
\hat{\varphi}_{\pm 1, 0, ... , 0} = \left(\frac{2}{h}\right)^{\frac{n}{4}}
\left(\frac{b_1 ... b_n}{a_1 ...  a_n}\right)^{\frac{1}{4}}
\exp\left( -\frac{\pi}{h}\sum_{i=1}^n \frac{b_i}{a_i}(q_i-x_i)^2\right), \eqno (A.10)
$$
where
$$ c_{\pm 1, 0, ... , 0} = \sqrt 2 \left(\frac{2}{h}\right)^{\frac{n}{4}}
\!\!\left(\frac{b_1 ... b_n}{a_1 ... a_n}\right)^{\frac{1}{4}}
\!\int\limits_{R^n} \hat{\varphi}_{\pm 1}\
\exp\left( -\frac{\pi}{h}\sum_{i=1}^n \frac{b_i}{a_i}(q_i-x_i)^2\right) dq.
$$

Now, let us return to the nonstationary  equations  (A.7).  Having
substituted in them the expressions  for $\hat{\varphi}_{k, k_1, ..., k_n}(q, x)$
 are  eigen
functions of the right hand side of the equation, we get equations
$$
\sum_{ k_1, ..., k_n = 0}^{\infty} \frac{\partial c_{k, k_1, ..., k_n}}{\partial \tau}
\hat{\varphi}_{k, k_1, ..., k_n}=
\sum_{ k_1, ..., k_n = 0}^{\infty} \lambda_{k, k_1, ..., k_n}\ c_{k, k_1, ..., k_n} \
\hat{\varphi}_{k, k_1, ..., k_n},
$$
  which due to orthogonality of the functions
$\hat{\varphi}_{k, k_1, ..., k_n}(q, x)$   in  the
space of square integrable over $q$ functions split into the  system
of equations
$$
 \frac{\partial c_{k, k_1, ..., k_n}}{\partial \tau} =
 \lambda_{k, k_1, ..., k_n}\ c_{k, k_1, ..., k_n}
$$
 for   $k\in Z\setminus \{0\}$, $ k_1 \geq 0, ... , k_n \geq 0.$
The solutions of these equations are of the form
$$
 c_{k, k_1, ..., k_n}(\tau, x, \xi) =  c_{k, k_1, ..., k_n}(0, x, \xi)
\exp(\lambda_{k, k_1, ..., k_n}\tau). \eqno   (A.11)
$$

Since $\lambda_{k, k_1, ..., k_n}<0$, then $c_{k, k_1, ..., k_n}$
            decrease exponentially  as
$\tau \to \infty$ and the largest contribution to
$\hat{\varphi}_k (\tau, q, x, \xi)$ is given for not
too small $\tau$ by the terms of the  series  (A.9)  with  the  largest
eigen value.

     Thus,  with  (A.9),  (A.11),  (A.8)  we  get  the   following
asymptotic in $\tau \to \infty$:
$$
\hat{\varphi}_k (\tau, q, x, \xi) \sim \frac{1}{\sqrt 2}c_{k, 0, ... , 0}(0, x, \xi)
\hat{\varphi}_{k, 0, ... , 0}(q, x)
\exp\left( -\tau\sum_{i=1}^n \frac{2\pi |k| a_i b_i}{h}\right).
$$

With (A.4) we deduce that in  (A.2)  every  summand $\varphi_k $  with
$k\ne 0$ exponentially decrease with the growth of $\tau$.
Hence  the  largest contribution to
$\varphi$ is given after a while by  the  terms  with  the
largest  exponent,  i.e.  for  $k = \pm 1.$  In  other  words,  we  have,
asymptotically,
$$\varphi \sim \varphi_{-1} + \varphi_1.$$

Having  substituted  here,  consecutively,  (A.4)  for  $k = \pm 1$,  the
asymptotic expressions for ¤«ï
$\hat{\varphi}_{\pm 1}$  obtained  above  and  expressions
(A.10) for $\hat{\varphi}_{\pm 1, 0, ... , 0}$ we finally get
\begin{eqnarray}
 \varphi(\tau, q, p, \xi) \!\!\!&\sim&\!\!\!
\frac{1}{\sqrt 2}\left( \frac{2}{h^3}\right)^{\frac{n}{4}}
\left(\frac{b_1 ... b_n}{a_1 ...  a_n}\right)^{\frac{1}{4}}
\exp\!\left( -\tau\sum_{i=1}^n \frac{2\pi |k| a_i b_i}{h}\right)
\times\nonumber\\
& &\times
\int_{R^n}
\exp\!\left( -\frac{\pi}{h}
\sum_{i=1}^n\frac{b_i}{a_i}(q_i-x_i)^2\right)
\times\nonumber\\
& &\times
\Biggl(
c_{-1, 0, ... , 0}(0, x, \xi)
\exp\!\left(-j\frac{2\pi k \langle p, x \rangle}{h}\right) +
\nonumber
\end{eqnarray}
$$\qquad \qquad \qquad \qquad
+c_{1, 0, ... , 0}(0, x, \xi)
\exp\!\left(j\frac{2\pi k \langle p, x \rangle}{h}\right)
\Biggr) dx, \eqno   (A.12)
$$
where $ c_{\pm 1, 0, ... , 0}(0, x, \xi) $ is the result of operations
$$
\varphi(q, p, \xi) = \varphi(0, q, p, \xi)
\mapsto \varphi_{\pm 1}(q, p, \xi)
\mapsto \hat{\varphi}_{\pm 1}(q, p, \xi)
\mapsto c_{\pm 1, 0, ... , 0}(0, x, \xi)
$$
according to the formulas (A.2), (A.5), (A.10).

     It is not difficult to verify that if $\varphi$ is  a  real  function

then
$ c_{ 1, 0, ... , 0}(0, x, \xi) = c^*_{-1, 0, ... , 0}(0, x, \xi)$.
Therefore, setting
$$\psi(x, \xi) = c_{-1, 0, ... , 0}(0, x, \xi)$$
 and having substituted this into  (A.12)
we get Theorem~3.1.

\newpage
\begin{center}{ \Large Appendix 2. Estimate of the parameter $a/b$ of the
considered model}\end{center}

 In this estimate we follow the method of [We] to justify  the
Lamb's shift. Consider the operator $A_f$
 given  by  (\ref{A_f_hidr})  for  the
Hamiltonian $f(q,p)$ as a perturbation of the  operator  $H$  for  the
hydrogen atom of the quantum mechanics.  The  perturbation  theory
implies  that  in  the  first  order  the  increment $\delta E_n$    of  the
eigenvalue $E_n$  of $H$ is of the form
$$\delta E_n= \int_{R^3}{\rho_n(x) (\bar V(x)-V(x) )dx},$$
where
$\rho_n(x)=|\psi_n(x)|^2$ and $\psi_n(x)$   is  the  eigenfunction  of  with
eigenvalue $E_n$  and where
$V(x)=-e^2/\sqrt{x_1^2+x_2^2+x_3^2 }=-e^2/r$.
 By  definition
the function $\bar{V}(x)$ is the mathematical  expectation
$\bar{V}(x)=M_q V(x+q)$,
where $q$ is normally distributed with density
$$\left(\frac{2b}{h a}\right)^{\frac{3}{2}} \exp\left(-\frac{2\pi  b}{h a}(q_1^2+q_2^2+q_3^2)\right). $$
After simplification we get:
                  $$\delta E_n =\int_{R^3} \bigl(M_q (\rho_n(x+q)) -\rho_n(x)\bigr) V(x) dx.$$

     Since $\rho_n(x)$ is smooth (infinitely  differentiable),  then  if
the standard deviation of $q$ is essentially smaller than the atom's
radius we can make use of the Taylor series expansion  of $ \rho_n(x+q)$
to compute $ M_q (\rho_n(x+q))$. We have
$$\rho_n(x+q) \approx (1+\langle q, \nabla\rangle+\frac{1}{2}\langle q, \nabla \rangle^2\rho_n(x).
$$
Since $M_q(q_i)=0,$ $M_q(q_i q_j)=0$ for $ i\ne j$
 and $M_q(q_i^2)=h a/(4\pi b),$ then we get
the approximate equality
$$ M_q (\rho_n(x+q))= \rho_n(x)+\frac {h a}{8\pi b}\Delta\rho_n(x),$$
where $\Delta ={\partial^2}/{\partial x_1^2}+
{\partial^2}/{\partial x_2^2}+{\partial^2}/{\partial x_3^2}$.

     Having substituted this $ M_q (\rho_n(x+q))$ into the  expression  for
$\delta E_n$  and since $\Delta $ is self adjoint, we get
      $$\delta E_n=  \frac{a h}{8\pi b}\int_{R^3}{\rho_n(x) \sum_{k=1}^3
\frac{\partial^2 V(x)}{\partial x_k^2}dx}.$$
Since for $V(x)=-{e^2}/{r}$   we have $\Delta V(x)=4 \pi e^2 {\delta}_0{(x)},$
 where ${\delta}_0{(x)}$  is
Dirac delta-function,
$$ \delta E_n=  \frac{a h e^2}{2 b} \rho_n(0). $$
For the hydrogen atom
$$\rho_n(0)=|\psi_n(0)|^2=\frac{1}{\pi n^3}\left(\frac{m e^2}{\hbar^2} \right)^3,$$
see e.g. [STZh] p.342, where $\hbar=h/(2\pi)$, and therefore
 $$ \delta E_n=  \frac{a\hbar \pi e^2}{b} \frac{1}{\pi n^3}\left(\frac{m e^2}{\hbar^2} \right)^3=
\frac{a}{b}\frac{m^3 e^8}{n^3 \hbar^5}=
\frac{a}{b}\frac{m^3 \alpha^4 c^4}{n^3 \hbar},$$
where $\alpha={e^2}/(\hbar c)={1}/{137}$ and  $c$ is the speed of light in vacuum. It follows
        $$\frac{a}{b}=\delta E_n\frac{n^3\hbar}{m^3 \alpha^4 c^4}.$$
In experiments of Lamb and Retherford [LR]it had been established that
for the hydrogen atom
$\delta E_2   = 1058 ŒgHz = 1058 \cdot 10^6 h$ erg.
Comparing this expression with our value of $\delta  E_2,$ we  directly  get
the estimate for $a/b$:
$$a/b  = 3,41\cdot 10^4 sec/ gr.$$
Accordingly, the standard deviation of $q$ in each coordinate is
   $$\Delta q_i=\sqrt{a\hbar/2b}=4,24\cdot 10^{-12}cm,$$
which is essentially smaller than the radius of the hydrogen atom.

\end{document}